\documentclass[draft]{agujournal2019}
\usepackage{url} 
\usepackage{lineno}
\usepackage[inline]{trackchanges} 
\usepackage{soul}
\usepackage{xcolor} 
\usepackage{amsmath}
\draftfalse

\journalname{Geophysical Research Letters}

\begin{document}

\title{Instability-Aware Steering of an Extreme Atmospheric River in an AI Weather Foundation Model}

\authors{
Moyan Liu\affil{1},
Qin Huang\affil{1},
and Upmanu Lall\affil{1,2}
}

\affiliation{1}{School of Complex Adaptive Systems \& Water Institute, Arizona State University, Tempe, AZ, USA.}
\affiliation{2}{Department of Earth and Environmental Engineering \& Columbia Water Center, Columbia University, New York, NY, USA.}

\correspondingauthor{Moyan Liu}{moyanliu@asu.edu}

\begin{keypoints}
\item Instability-aware perturbation sites are identified using FTLE and jet-eddy interaction criteria in an AI weather foundation model
\item Idealized thermodynamic perturbations produce nonlinear downstream changes in atmospheric river moisture transport
\item Perturbation outcomes vary with the local synoptic kinematic environment at the seeding location
\end{keypoints}

\begin{abstract}
Advances in Deep Learning methods for weather forecasting are creating the opportunity to computationally explore the potential for the steering or control of extreme weather trajectories for societal risk reduction. We present initial investigations into the feasibility of redirecting extreme atmospheric rivers (ARs) through small, instability-aware perturbations. Using the Aurora AI weather foundation model, we identify sensitive upstream locations using finite-time Lyapunov exponents and jet-eddy interaction criteria. We apply an idealized cloud-seeding operator that mimics latent heat release to assess whether these Lyapunov-guided interventions can influence downstream evolution. In a case study of a severe California AR, perturbations induce coherent downstream shifts in moisture transport, reducing intensity at landfall under favorable kinematic conditions. The response is nonlinear and contingent on the local flow geometry. These initial results suggest that the atmosphere's intrinsic chaotic sensitivity could be leveraged for dynamical control, offering a new research direction for extreme event risk mitigation.
\end{abstract}

\section*{Plain Language Summary}
Extreme weather events, such as the ``atmospheric rivers'' cause severe flooding. They are becoming more frequent and intense due to climate change. Modern AI weather models have become quite good at predicting these events. We explore whether the atmosphere’s natural ``chaos,'' i.e., its sensitivity to small changes, could be used to subtly nudge such an extreme event off its destructive path. Using Aurora, often considered one of the best-performing AI weather foundation models, we identified "sensitive spots" where the atmosphere is most suitable for propagating a perturbation. We consider a cloud-seeding mechanism to induce a small perturbation in temperature and humidity at such candidate locations. Our results show that under the right conditions, these tiny upstream adjustments can significantly reduce the intensity of an atmospheric river by the time it reaches land. This opens the possibility that we could go beyond predicting disasters to one day using the atmosphere’s own complexity to mitigate the risks of extreme weather.

\section{Introduction}
Anthropogenic climate change is amplifying the frequency, intensity, and persistence of extreme weather events, including heatwaves, floods, tropical cyclones, and atmospheric rivers \cite{Clarke_2022,Pradhan_2022}. These events increasingly dominate climate-related impacts, threatening infrastructure \cite{Hwang_2024}, economies \cite{Newman_2023,Ferreira_2024}, and human safety \cite{Bell2018}. Many such extremes arise from nonlinear amplification within dynamically organized atmospheric structures (e.g. Rossby waves and Baroclinic waves) \cite{Lembo2024}. These systems exhibit strong transient growth and finite-time instability, allowing localized perturbations to propagate and reorganize large-scale circulation patterns. As climate variability intensifies, understanding how such dynamical structures evolve and whether their trajectories can be influenced becomes important for managing the risks associated with high-impact weather extremes.

Over the past few decades, substantial progress has been made in weather and climate prediction, leading to improved forecast skill and longer lead times for many extreme events \cite{Vitart2018, Domeisen2022}. However, strategies for using these forecasts to influence the evolution or severity of extreme events remain largely unexplored. This growing gap between predictive capability and actionable intervention motivates the exploration of complementary, intervention-oriented frameworks.

We explore an application of the Weather Jiu-Jitsu (WJJ) framework, which leverages the intrinsic chaotic dynamics of weather systems to subtly redirect or dissipate their destructive trajectories through precisely timed, small-energy interventions \cite{Huang2025WeatherJiuJitsu}. Proof-of-concept studies in idealized, low-dimensional chaotic systems have shown that control strategies based on local dynamical instability, rather than full state predictability, can influence system trajectories even in the presence of noise and limited observations \cite{Liu2025TargetedAdaptiveChaos,Liu2025RegimeIdentificationControl}. However, whether such concepts can be extended to realistic, high-dimensional atmospheric systems remains an open and central challenge.

Recent advances in AI and foundation weather models produce fast, high-resolution forecasts at a fraction of the computational cost of traditional numerical weather prediction systems \cite{Das2024HybridPhysicsAI,Price2025ProbabilisticWeatherML,Lang2024AIFS}. They provide the potential for systematic perturbation experiments. However, extending instability-aware control strategies from low-dimensional Lorenz systems to high-dimensional atmospheric systems is challenging because instability geometry may differ across models and reanalysis products due to data assimilation uncertainty and structural differences in the underlying dynamics. A necessary first step, therefore, is to establish whether instability-aware perturbations produce coherent responses within a single high-dimensional model before examining their robustness across models or real-world applications.

We select Atmospheric Rivers (ARs) as an initial test case for instability-aware perturbation experiments. ARs are dynamically organized moisture transport structures that emerge from nonlinear interactions between the midlatitude jet stream and baroclinic eddies. In this respect, they represent a high-dimensional analog of the jet-eddy instability mechanisms captured in idealized Lorenz models \cite{lorenz1984}. Just as Lorenz systems exhibit regime transitions driven by nonlinear energy exchange between mean flow and eddies, AR evolution depends sensitively on jet positioning, deformation fields, and moisture transport pathways. This dynamical structure makes ARs a natural candidate for testing whether chaos-control principles developed in low-dimensional settings retain their leverage in distributed atmospheric systems \cite{Liu2025TargetedAdaptiveChaos}. At the same time, ARs are a leading driver of extreme precipitation and coastal flooding, ensuring that the analysis is not only dynamically informative but societally relevant.

Here, we take a first step toward translating WJJ control concepts to realistic weather systems by applying targeted perturbations within an AI foundation weather model, Aurora, that has been shown to be one of the best at forecasting weather extremes. Motivated by the physical mechanism of cloud seeding, we investigate whether and under which conditions small, localized perturbations of atmospheric states can influence the evolution of extreme weather trajectories. Our aim is not to propose operational weather control, but rather to assess the feasibility, sensitivity, and chaos-aware intervention strategies in a realistic atmospheric model. 

While prior studies have explored perturbation-based approaches in numerical models to amplify extremes and uncover plausible but unobserved high-impact scenarios under climate change \cite{Justin2026}. Our objective differs fundamentally. Instead of seeking to expand the envelope of possible extremes, we focus on examining both their potential to amplify and to suppress extreme events within the model dynamics. This perspective allows us to frame perturbations as probes of dynamical sensitivity and controllability. To our knowledge, AI based weather forecast foundation models have not previously been used in this way.

\section{Methodology}
First, we introduce the Aurora weather foundation model, which serves as the forecasting backbone for all experiments. Second, we identify dynamically sensitive perturbation locations using the finite-time Lyapunov exponent (FTLE) in combination with physical constraints related to jet–eddy interactions and moisture transport. Finally, we introduce an idealized cloud-seeding based perturbation operator that represents localized latent heat release and moisture removal within the model.

\subsection{Aurora Weather Foundation Model}

Aurora is a large-scale AI weather foundation model designed to produce fast, high-resolution forecasts across a wide range of atmospheric variables \cite{Bodnar_2025_aurora}. The model is trained on multi-decadal atmospheric datasets spanning historical records through 2020, including ten reanalysis, forecast and simulation datasets. Aurora employs a hybrid architecture combining a 3D Swin Transformer U-Net backbone with Perceiver encoders and decoders, enabling the model to capture complex multiscale atmospheric dynamics. Benchmark studies have shown that Aurora achieves competitive or superior skill in predicting high-impact weather extremes \cite{Zhao2025ExEBench}, including extreme precipitation \cite{Barriopedro2025}, tropical cyclones \cite{Sahu2025}, and atmospheric rivers \cite{huangliu2026}.

In this study, Aurora is employed in a feed-forward forecasting configuration in which localized perturbations are introduced at selected forecast times and locations, after which the atmospheric state is integrated forward in time. Both perturbed and unperturbed forecasts are initialized from identical baseline states, allowing the effects of the perturbations to be isolated and enabling direct comparison of trajectory divergence, event intensity, and the spatial evolution of the flow. We use the pretrained Aurora 0.25° global model, which operates on a horizontal grid of 721 × 1440 points and produces forecasts at 6-hour intervals (00, 06, 12, and 18 UTC). All experiments use data outside Aurora's training period, i.e., post 2020.

\subsection{Perturbation Site Selection}
To identify locations where small perturbations are most likely to trigger disproportionate downstream responses, we employ diagnostics based on Finite Time Lyapunov Exponents \cite{Garaboa-Paz_2017,Perez2021} (FTLE). FTLE fields quantify the rate of separation between neighboring trajectories over a finite time window and thus provide a measure of local flow instability and sensitivity to initial conditions. Regions of elevated FTLE indicate strong stretching and deformation in the flow and are therefore natural candidates where small perturbations may amplify rapidly.

To quantify dynamical sensitivity in the flow, we compute FTLE fields along the trajectories of Lagrangian tracers at 850 hpa, where the low-level jet core responsible for AR moisture transport is located. Following \cite{Garaboa-Paz_2017},  the FTLE at an initial position $\mathbf{r}_0$,
reference time $t_0$, and integration time $\tau$ is defined as
\begin{equation}
\lambda(\tau,t_0,\mathbf{r}_0)
= \frac{1}{|\tau|}\,
\log\!\left(
\sqrt{\mu_{\max}\!\left(\tilde{\mathbf{C}}(\tau,t_0,\mathbf{r}_0)\right)}
\right),
\end{equation}
where $\mu_{\max}$ denotes the maximum eigenvalue of the pull-back Cauchy-Green deformation tensor $\tilde{\mathbf{C}}$.
The deformation tensor is given by
\begin{equation}
\tilde{\mathbf{C}}(\tau,t_0,\mathbf{r}_0)
=
\left(
\nabla \mathbf{r}(t_0+\tau; t_0,\mathbf{r}_0)
\right)^{\!T}
\mathbf{G}(\theta(\tau))
\nabla \mathbf{r}(t_0+\tau; t_0,\mathbf{r}_0),
\end{equation}
where $\nabla \mathbf{r}$ is the gradient of the flow map and
$\mathbf{G}$ accounts for the metric on the spherical domain with $\theta(\tau)$ denoting the latitude of the particle at the final time. Forward FTLE fields are computed by integrating Lagrangian trajectories using the horizontal wind fields $(u, v)$ at 850 hPa produced by the Aurora model over a prescribed integration horizon.

While FTLE identifies dynamically sensitive regions, not all such locations are meteorologically relevant for influencing atmospheric river evolution. Therefore, we apply additional physical constraints to isolate perturbation sites that are both dynamically unstable and physically useful. Specifically, candidate locations are required to satisfy the following criteria:

\begin{itemize}
    \item First, candidates are selected from regions exhibiting persistently high FTLE values (top 10\% of the distribution), as these locations are expected to amplify small perturbations most efficiently.
    \item Second, candidates must lie within or adjacent to the ARs, defined as regions where the integrated vapor transport (IVT) exceeds $250\ \mathrm{kg\,m^{-1}\,s^{-1}}$. This ensures that perturbations act on the moisture transport pathway rather than on sensitive flow structures far removed from the ARs. 
    \item Third, we require candidates to lie near the boundary of the Lagrangian reachability set. This set is computed via forward trajectory integration and identifies air parcels that reach the target area within the forecast horizon. 
    \item Fourth, to ensure physical relevance at the steering level, candidates are further filtered using upper tropospheric jet criteria. Specifically, we retain only locations where the 250 hPa horizontal wind speed satisfies $30\ \mathrm{m\,s^{-1}} < U(x,y) < 60\ \mathrm{m\,s^{-1}}$, corresponding to the flanks of the jet stream rather than the jet core itself. These regions play an important role in jet stream and eddy interactions, where synoptic scale disturbances exchange energy with the mean flow and strongly influence the growth, steering, and amplification of extreme weather systems. 
    \item Fifth, all candidates within the target area are excluded, ensuring that perturbations are applied upstream of the landfall zone where they have sufficient lead time to propagate through the flow dynamics and \item Sixth, perturbations are restricted to grid points and vertical levels where $\mathrm{RH} > 0.8$, ensuring that interventions are applied only where moist processes are active.
    \item Finally, to avoid clustering perturbations within a single localized structure and to ensure spatial independence among selected sites, a minimum separation distance of 500 km is enforced between candidates. 

\end{itemize}

This multi-criteria framework ensures that each perturbation site has a physically grounded mechanism through which it can influence the downstream extreme event.

\subsection{Thermodynamic Intervention}
We consider an idealized cloud-seeding based perturbation strategy because latent heat release associated with phase changes has been shown to influence large-scale atmospheric circulation. Recent studies show that diabatic heating from condensation and deposition can strengthen subtropical jets, modify upper tropospheric waveguides, and destabilize the Hadley cell structure, thereby exerting a nonlocal influence on synoptic and planetary-scale flow evolution \cite{Lin2025HadleyCellInstability,Auestad2025}. These processes make localized latent heating of condensation a physically meaningful and dynamically efficient lever for perturbing atmospheric trajectories.

Aurora does not explicitly represent cloud microphysics or chemical seeding agents. Direct simulation of operational cloud-seeding processes is therefore not possible within the model. Instead, we implement an idealized cloud-seeding operator that mimics the essential thermodynamic consequences of cloud seeding, while remaining consistent with the prognostic variables available in the model. Precipitation generated by seeding is diagnosed as a fraction of humidity that is converted to frozen condensate and allowed to sediment out of the atmospheric column. The associated latent heat release from vapor deposition is computed consistently and added to the temperature field, ensuring thermodynamic coherence between moisture removal, precipitation production, and atmospheric warming. This approach enables physically interpretable perturbations without introducing explicit microphysical or chemical processes.

Seeding is restricted to a subset of the grid points within the prescribed spatial mask defined by the site selection procedure and to vertical levels where the local relative humidity exceeds a threshold, $\mathrm{RH} > 0.8$. In active grid cells, a fraction of the local specific humidity is converted to ice, parameterized as
$q_{\mathrm{frozen}}=\min\left(\eta\,q,\;f_{\max}\,q\right)$ where $\eta$ is the freezing efficiency and $f_{\max}$ is a hard upper bound on moisture removal. The removal of water vapor is accompanied by the release of latent heat of deposition, such that the temperature tendency is given by
\begin{equation}
\Delta T = \frac{L_d}{c_p}\, q_{\mathrm{frozen}},
\end{equation}
where $L_d = L_v + L_f$ is the latent heat of deposition, with $L_v$ the latent heat of vaporization and $L_f$ the latent heat of fusion, and $c_p$ is the specific heat of air at constant pressure. Temperature and moisture are updated at each target pressure levels ($p_i$) simultaneously as $T'(p_i)=T(p_i)+\Delta T(p_i)$ and $q'(p_i)=q(p_i)-q_{\mathrm{frozen}}(p_i)$, enforcing local consistency between moisture removal and latent heating. A diagnostic fraction of the frozen condensate is assumed to precipitate out: $q_{\mathrm{precip}}(p_i)=f_{\mathrm{fallout}}\,q_{\mathrm{frozen}}(p_i)$, without explicitly resolving cloud microphysical processes. Because these properties vary with pressure, the seeding perturbation is applied independently at each seeded level and the overall atmospheric response represents the vertically integrated effect across the seeded layers. This scheme provides a controlled, physically interpretable representation of humidity removal and latent heating suitable for sensitivity experiments and instability-aware perturbation studies.

\section{Results}
We implement this procedure for a severe atmospheric river event impacting the U.S. West Coast on 27 December, 2022. This event ranks among the most intense ARs observed along the coast in recent decades and was characterized by strong moisture transport and widespread hydrometeorological impacts. Integrated vapor transport (IVT) exceeded $1250~\mathrm{kg\,m^{-1}\,s^{-1}}$ over large portions of coastal and interior California, far surpassing commonly used thresholds for extreme AR classification \cite{DeFlorio2024CaliforniaDroughtFlooding}. The event produced severe flooding and power outages, making it a suitable test case for assessing whether small, physically consistent interventions can meaningfully alter trajectory. As such, it provides a demanding benchmark for evaluating the sensitivity and dynamical response of the proposed perturbation framework.

\subsection{Site Selection Result}
Figure 1 illustrates the outcome of the perturbation site selection procedure applied to the Aurora forecast fields. The background shows regions of FTLE value, while the overlaid dashed contours indicate the upper-tropospheric jet stream. The selected perturbation sites (yellow markers) are a sparse set of locations situated along the flanks of the jet, where dynamically sensitive flow structures intersect with physically relevant circulation features.

The spatial distribution of the perturbation sites reflects a balance between dynamical leverage and physical relevance. Sites are sufficiently separated to avoid clustering within a single coherent structure, while remaining aligned along the large-scale waveguide of the jet. These results demonstrate that the site selection procedure yields a limited number of target perturbation locations that satisfy both objective dynamical criteria and physically informed circulation constraints. These sites form the basis for the cloud seeding based perturbation experiments presented in the subsequent section.

\begin{figure}[!htbp]
  \includegraphics[width=0.95\textwidth]{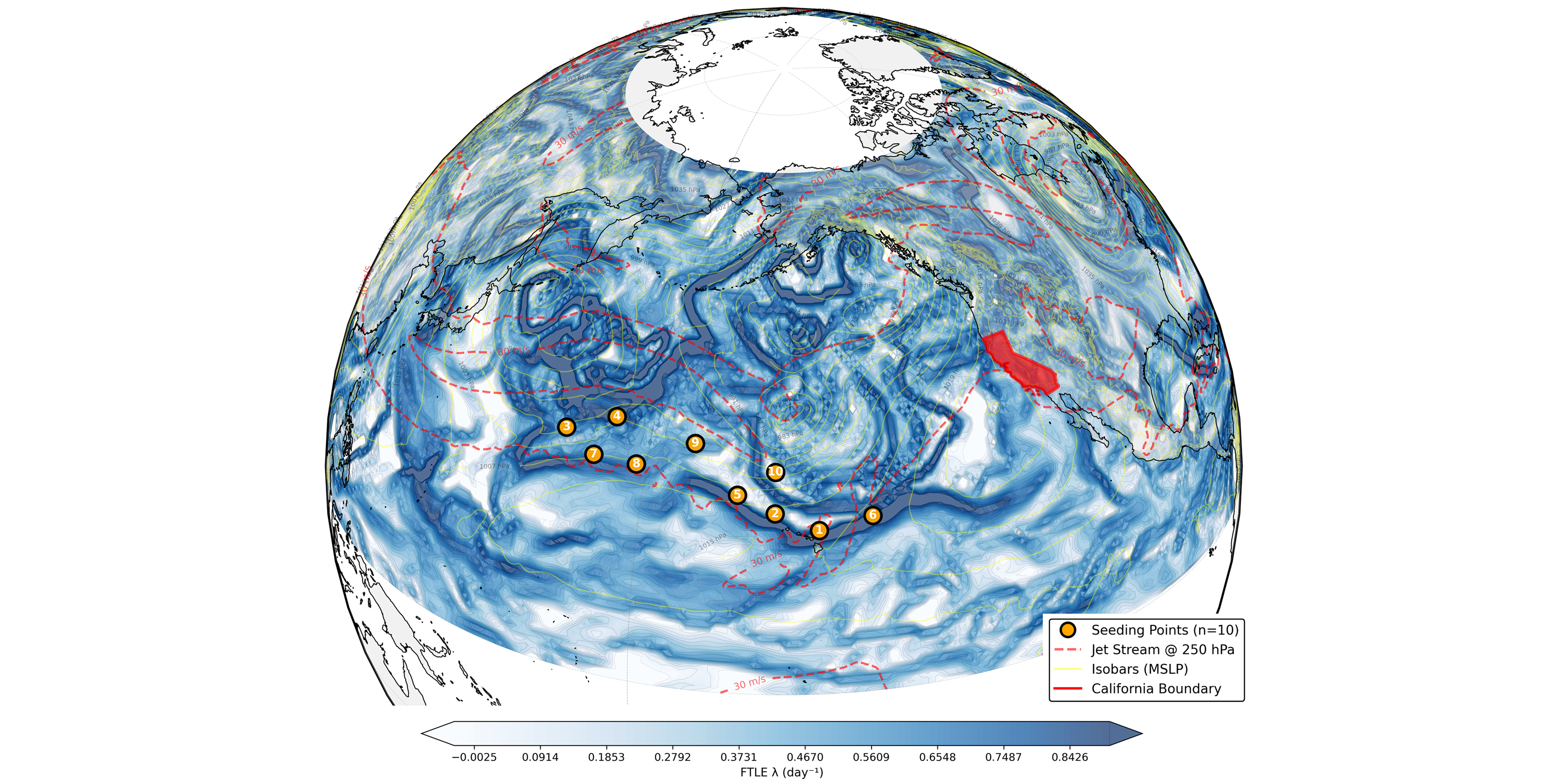}
  \caption{Selected Perturbation Sites for Atmospheric River Intervention}
  \label{fig:ftle}
\end{figure}

\subsection{Perturbations Result}
Each candidate perturbation is applied once over a single 6-hour model step within a 300 km radius, targeting the lower-to-mid troposphere ($925$--$700$~hPa). In this configuration, an idealized 30\% freeze efficiency is applied to the local vapor field, with 50\% of the frozen condensate removed as precipitation. Within the seeded regions, the conversion ranges from approximately $2.3$--$3.4\,\mathrm{g\,kg^{-1}}$, producing local temperature increases of about $6.7$-$10.4~\mathrm{K}$ through latent heat release. Because the perturbation is applied only once over a 6-hour interval, the resulting temperature and moisture adjustments remain localized relative to the background variability of the atmospheric river environment. The imposed heating and drying therefore act as a spatially and temporally limited thermodynamic adjustment rather than large-scale forcing.

The response to localized perturbations exhibits a nonlinear character, with small initial heating anomalies propagating forward in time and reorganizing the large-scale flow geometry. This nonlinear evolution ultimately reveals changes to the Integrated Vapor Transport (IVT), which quantifies the vertically integrated horizontal transport of water vapor and is widely used to characterize atmospheric rivers. Table 1 summarizes all selected perturbation sites, including the selection criteria and the resulting fractional change in IVT at the target region.
\begin{table}
\caption{Summary of perturbation site selection ranked by FTLE. Positive $\Delta$IVT indicates reduction in integrated vapor transport.}
\centering
\begin{tabular}{l c c c c c c c c}
\hline
\# & Lat ($^\circ$N) & Lon ($^\circ$E) & FTLE & $\Delta$IVT (\%) & Conv $\times 10^{5}$ & Vort $\times 10^{5}$ & Strain $\times 10^{5}$ \\
\hline
1  & 21.8 & 204.8 & 1.0548 & $-0.09$ & 0.32  & $-5.43$ & 1.63 \\
2  & 24.0 & 198.8 & 0.9916 & 0.32    & 4.48  & $-2.30$ & 7.03 \\
3  & 30.8 & 166.5 & 0.9415 & \textbf{\textcolor{blue}{5.21}} & 2.54  & 9.64    & 9.82 \\
4  & 33.8 & 174.0 & 0.9319 & 0.67    & $-0.42$ & $-4.34$ & 5.54 \\
5  & 26.2 & 193.5 & 0.9306 & $-0.89$ & $-0.65$ & $-2.96$ & 2.69 \\
6  & 23.2 & 212.2 & 0.8913 & $-0.20$ & 0.24  & $-2.35$ & 3.78 \\
7  & 28.5 & 171.8 & 0.8192 & \textbf{\textcolor{red}{-1.39}} & 2.38  & $-3.23$ & 3.89 \\
8  & 28.5 & 178.5 & 0.7927 & 0.15    & $-0.43$ & $-3.38$ & 1.77 \\
9  & 32.2 & 186.8 & 0.6910 & 0.49    & 2.48  & $-1.26$ & 4.17 \\
10 & 29.2 & 198.8 & 0.5723 & 0.25    & 6.11  & 8.22    & 10.91 \\
\hline
\multicolumn{9}{l}{$^{*}$Conv $>0$: convergent, $<0$: divergent; Vorticity $>0$: cyclonic, $<0$: anticyclonic; Strain$\ge 0$}
\end{tabular}
\end{table}

Importantly, the sign and magnitude of $\Delta IVT$ vary across sites despite comparable perturbation magnitudes, indicating that the system response depends sensitively on the local kinematic structure of the flow rather than on forcing amplitude alone. Positive $\Delta IVT$ values indicate steering or weakening of the ARs, whereas negative values correspond to amplification. The coexistence of both outcomes confirms that the IVT response emerges from nonlinear interactions within the evolving synoptic-scale circulation.

\subsection{Case Study Result}
\subsubsection{Synoptic Evolution}

To examine the dynamical mechanisms underlying the bifurcated IVT responses, we analyze two representative perturbation experiments: a “good” case (\#3) yielding a downstream IVT reduction and a “bad” case (\#7) producing amplified IVT. Both perturbations are applied at the same configuration. The contrasting outcomes therefore isolate the role of local flow geometry rather than forcing amplitude.

In the “good” case shown in figure 2 (top), the perturbation introduces a coherent deformation of the IVT filament, displacing the moisture plume poleward. The IVT anomaly develops into a spatially organized dipole structure consistent with trajectory deflection. By contrast, in the bad case at figure 2 (down), the initial anomaly remains embedded within the jet core, and subsequent evolution shows reinforcement of the pre-existing moisture transport rather than filament displacement. The anomaly structure becomes more axisymmetric and aligned with the jet, leading to net IVT enhancement in the target region.

\begin{figure}[!htbp]
  \includegraphics[width=\textwidth]{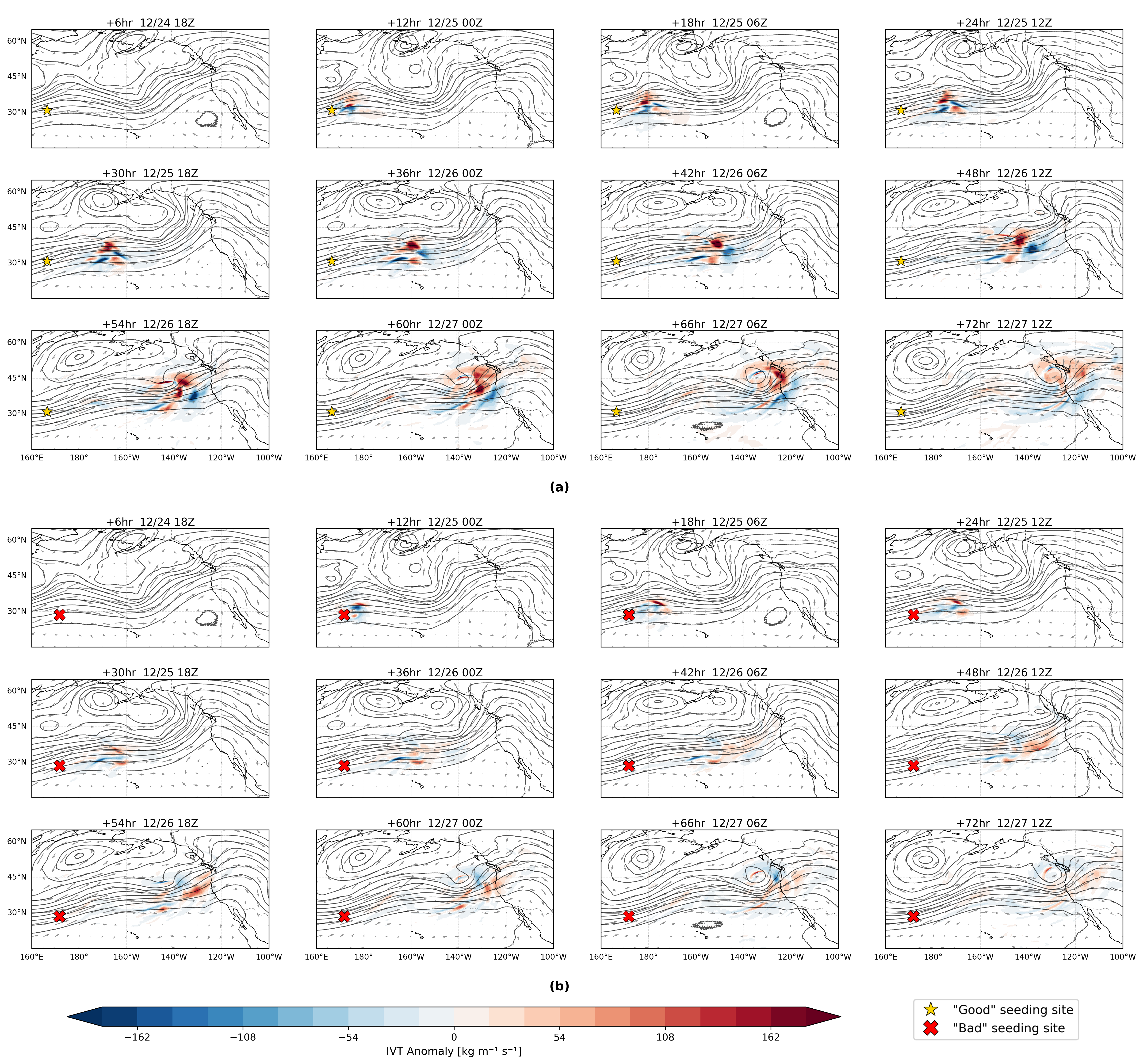}
  \caption{The time evolution of Z500, 500~hPa wind, and IVT anomaly for two seeding experiments. (a) \textit{Favorable site} ($\star$, 30.8$^\circ$N, 166.5$^\circ$E): the seeding location lies within a cyclonically shearing, jet-adjacent flow regime with high FTLE. The imposed perturbation is amplified through the flow, leading to a reduction of IVT in the California target region during landfall. (b) \textit{Unfavorable site} ($\times$, 28.5$^\circ$N, 171.8$^\circ$E): the seeding location is situated within an anticyclonically shearing environment. Despite identical perturbation magnitude, it leads to an increase in IVT over the target region.}
  \label{fig:2}
\end{figure}

To quantify the local flow environment at each seeding location, we computed 850~hPa kinematic diagnostics at the moment of perturbation, following the kinematic decomposition framework of \cite{Dacre2025}. The horizontal flow at each candidate site was characterized by three quantities: convergence (Conv), relative vorticity ($\zeta$), and total deformation strain ($S$), evaluated via central finite differences on the $0.25^\circ$ grid (Table~1).
\begin{align}
\text{Conv} &= -\left(\frac{\partial u}{\partial x} + \frac{\partial v}{\partial y}\right), \\
\zeta &= \frac{\partial v}{\partial x} - \frac{\partial u}{\partial y}, \\
S &= \sqrt{\left(\frac{\partial u}{\partial x} - \frac{\partial v}{\partial y}\right)^2 + \left(\frac{\partial v}{\partial x} + \frac{\partial u}{\partial y}\right)^2}.
\end{align}

The contrasting behavior between the two cases can be interpreted through this kinematic framework. The “good” case is located in a strain-dominated, cyclonically sheared environment. The perturbation projects onto a region of enhanced horizontal shear along the periphery of a synoptic-scale vortex. The localized diabatic heating modifies the deformation field, increasing strain and inducing Lagrangian trajectory separation. This results in lateral displacement of the moisture filament and a downstream weakening of IVT at landfall. The “bad” case, by contrast, exhibits strain with anticyclonic background vorticity. Here, rather than inducing trajectory separation, the perturbation maintains alignment with the mean advective flow. The anticyclonic environment rotates the moisture anomaly back into the jet axis rather than deflecting it poleward, leading to enhanced moisture flux convergence and increased IVT at the coast.

Importantly, IVT response depends sensitively on initial phase space location of the perturbation within the evolving synoptic flow. The divergence between two cases emerges progressively with lead time, reflecting nonlinear moisture and vorticity coupling. The coexistence of reduction and amplification outcomes indicates that ARs sensitivity is intrinsically nonlinear and phase dependent. Rather than prescribing a deterministic effect of cloud seeding, the results show that the impact of perturbation is contingent upon its projection onto the local deformation tensor of the flow. It suggests that effective perturbation strategies must account for the evolving kinematic structure of the baroclinic wave, particularly the spatial distribution of strain and convergence.

\subsubsection{Temporal and Spatial IVT Response at Target Region}
To quantify the downstream impact of the kinematic deformation described above, Figure 3 shows the temporal evolution and spatial structure of IVT anomalies over the California region. The top panel presents the IVT time series for ERA5 ground truth, Aurora control, and Aurora perturbed simulations. During the landfall window (the shaded region), the perturbed simulation exhibits a reduction, showing that the kinematic deformation described above translates into a downstream weakening of moisture transport. However, the model forecasts underestimate the observed peak IVT from ERA5, and this gap highlights the need for future work on accurate representation of extreme ARs events. 

The middle row quantifies the fractional IVT change for ARs threshold ($250\ \mathrm{kg\,m^{-1}\,s^{-1}}$) at event windows. Generally at a normal ARs threshold, the perturbation produces a spatial suppression of IVT within the target area. The largest suppression occurs at +66 h, coincident with peak deformation of the filament as it approaches the coast. The persistence of positive mean reduction across all three lead times confirms that the upstream kinematic modification systematically weakens moisture transport within the broader ARs envelope. The bottom row isolates the extreme ARs core ($800\ \mathrm{kg\,m^{-1}\,s^{-1}}$), which represents the most intense moisture transport and is typically associated with peak precipitation impacts. The extreme core experiences a suppression at event peak hour. 

However, the present analysis considers only a single perturbation applied at one location and one time, with evaluation focused on a single target region. In realistic scenarios, any intervention strategy would require a more sophisticated framework incorporating multiple spatial and temporal constraints. Effective implementation would likely involve coordinated multi-stage perturbations applied at different locations and forecast times, informed by evolving flow geometry and competing objectives. Learning the features that would be most promising a priori, and integrating that knowledge into a control strategy is an open research question.

\begin{figure}[!htbp]
  \includegraphics[width=\textwidth]{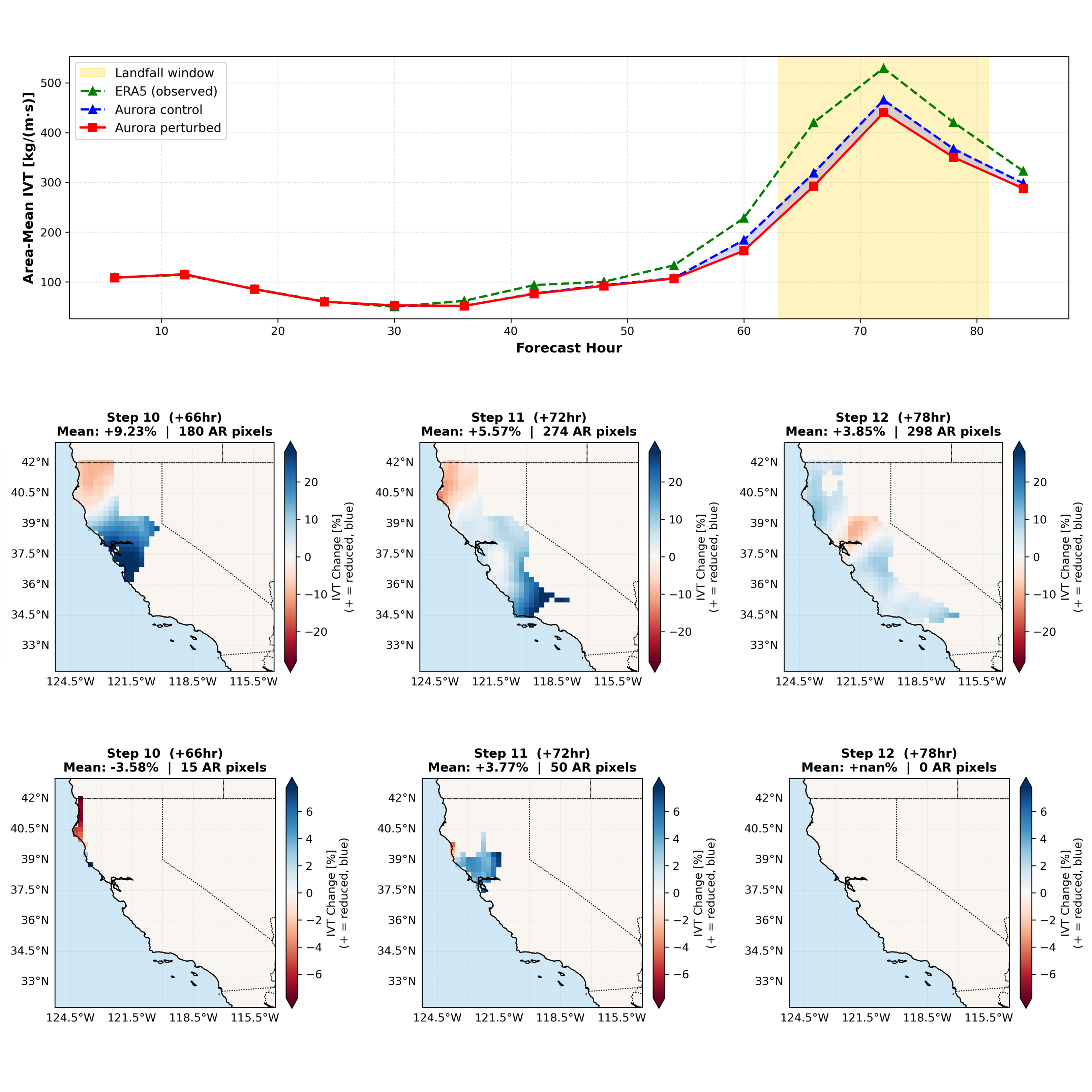}
  \caption{Time evolution and spatial structure of IVT response to targeted perturbations during an atmospheric river event. Top panel: Area-mean IVT over the target region for ERA5 (observed), Aurora control, and Aurora perturbed simulations, with the shaded region indicating the landfall window. Middle row: Percent change in IVT for AR pixels ($250\ \mathrm{kg\,m^{-1}\,s^{-1}}$) at selected forecast steps. Bottom row: Percent change in IVT for extreme AR core pixels ($250\ \mathrm{kg\,m^{-1}\,s^{-1}}$). Positive values indicate IVT reduction relative to control (blue color).}
  \label{fig:3}
\end{figure}

\section{Discussion and Conclusions}
Our results show that extreme-event trajectories within a modern AI weather foundation model contain structured finite-time instability geometries that can be exploited to alter downstream impacts. By combining dynamical sensitivity diagnostics with physically motivated perturbations, we demonstrate that localized thermodynamic adjustments can induce coherent downstream changes in moisture transport and landfall intensity. The resulting responses arise not from direct forcing of the extreme event itself, but from leveraging intrinsic atmospheric sensitivity upstream, consistent with the Weather Jiu-Jitsu framework.

A key scientific finding is that the effectiveness of perturbations depends strongly on the local dynamical geometry of the flow. In particular, perturbations introduced within regions characterized by cyclonic shear and strong deformation are more likely to induce downstream displacement and weakening of the moisture filament, whereas perturbations aligned with anticyclonic or advective flow structures tend to reinforce moisture convergence and maintain trajectory coherence. This asymmetry highlights the role of deformation and shear in controlling Lagrangian trajectory separation and provides a physically interpretable mechanism linking FTLE diagnostics to downstream impact.

These behaviors are consistent with classical baroclinic instability theory, in which regions of enhanced shear and deformation promote the growth and reorganization of synoptic structures. In this sense, the sensitivity patterns identified here are not artifacts of the AI model, but reflect underlying dynamical processes governing atmospheric transport. The present framework therefore provides a bridge between data-driven weather prediction and established dynamical systems understanding of atmospheric flow.

More broadly, the experiments suggest a conceptual distinction between predictability limits and controllability potential in atmospheric systems. Even in flows where forecast uncertainty grows rapidly, structured instability may create windows in which small, targeted perturbations can influence trajectory evolution before nonlinear amplification fully develops. Chaotic sensitivity may therefore represent not only a barrier to prediction but also a potential source of dynamical leverage.

Several limitations must be acknowledged. First, the perturbations considered here are applied as a single intervention lasting one model time step (6 hours). In practice, any realistic strategy would likely require adaptive or repeated perturbations that respond to the evolving flow. Second, the cloud-seeding based operator is highly idealized and represents bulk thermodynamic effects rather than detailed microphysical processes. This study should therefore not be interpreted as proposing an operational weather modification strategy. Rather, it demonstrates the existence of dynamically responsive regimes in which localized perturbations can influence atmospheric evolution within a controlled modeling environment.

In addition, the present analysis focuses on a single atmospheric river event. This choice allows a detailed, process-level examination of the mechanisms through which perturbations interact with flow geometry, but does not address robustness across events or initial condition uncertainty. Future work will extend this framework using ensemble-based approaches, incorporating stochastic perturbations and multiple case studies. By demonstrating that modern AI weather models contain dynamically exploitable instability structures, this work provides the first example of instability-aware perturbation design in a high-dimensional forecasting system. While exploratory, the results suggest a new research direction at the intersection of nonlinear atmospheric dynamics, machine learning, and extreme event risk management.

Finally, we emphasize that any practical implementation of such ideas would raise substantial ethical, environmental, and equity considerations, including the potential displacement of impacts across regions. These questions are beyond the scope of the present study, which focuses solely on identifying and understanding dynamical sensitivity in atmospheric flows.

%
%

\section*{Open Research Section}
The ERA5 reanalysis datasets used in this study are publicly available from the Copernicus Climate Data Store: ERA5 hourly data on single levels from 1940 to present (\url{https://doi.org/10.24381/cds.adbb2d47}) and ERA5 hourly data on pressure levels from 1940 to present (\url{https://doi.org/10.24381/cds.bd0915c6}). The Aurora foundation model is open-source and available via the Microsoft Aurora GitHub repository (\url{https://github.com/microsoft/aurora/}).

\section*{Conflict of Interest declaration}
The authors declare that there are no conflicts of interest for this manuscript.

\acknowledgments
The authors declare that no funds, grants, or other support was received during the preparation of this manuscript.

\bibliography{reference}

\end{document}